\documentclass[fleqn,twoside]{article}
\usepackage{amsmath,espcrc2,graphicx}

\usepackage{graphicx}
\usepackage[figuresright]{rotating}

\title{An experimental foundation for electromagnetic shower formation in the geomagnetic field}
\author{U.I. Uggerh{\o}j\address{Department of Physics and Astronomy, University of Aarhus, Denmark}}

\begin{document}






\begin{abstract}
At very high energies the Earth magnetic field, although in many other connections considered rather weak, acts as a strong field on photons and leptons. This paper discusses the intimate connection between this effect and the corresponding `strong field effects' observed at accessible accelerator energies in aligned single crystals. As such, these effects constitute an experimental verification of the theoretical basis used for simulations of the development of electromagnetic showers in magnetic fields, in particular the geomagnetic field. A short discussion of more general aspects of the shower development in the fields present at different distance scales is included.
\vspace{1pc}
\end{abstract}

\maketitle


\section{Introduction}

With the advent of facilities like the Pierre Auger Observatory \cite{Wats98} for the detection of ultra high energy cosmic rays of energies in the EeV ($10^{18}$ eV) region and orders of magnitude above, pair production and photon emission in the magnetic field of the Earth become increasingly relevant. These phenomena have been studied as early as the late thirties by Pomeranchuk \cite{Pome40}, later by McBreen and Lambert in the early eighties \cite{McBr81} and recently e.g.\ by Stanev and Vankov \cite{Stan98} as well as in extended air shower (EAS) simulations by Plyasheshnikov and Aharonian \cite{Plya02}. An important issue in this context is the possibility of distinguishing photon-initiated EASs from those initiated by protons or heavy nuclei as this distinction may shed light on the question of `top-down' (topological defects, massive X-particles) or `bottom-up' (acceleration of known particles) mechanisms \cite{Shin02,Stan98b}. Another issue is the directional dependence of photon-initiated horizontal showers as opposed to neutrino-initiated showers that bear no evidence of the magnetic field \cite{Cape98,Bert00}.

The aim of this paper is primarily to discuss the experimental justification for the phenomenon and secondly to address the question of `strong field' shower development in the many different magnetic fields which may be encountered by a photon of extragalactic origin. 

\section{Pair production in a magnetic field}

In the following, pair production and radiation emission processes are considered as taking place in magnetic or electric fields that are homogenous (constant in space over the formation length of the corresponding process).
The pair production probability differential in the energy of one of the final state particles, $\epsilon$, is given as \cite{Baie98}
\begin{equation}
\frac{\mathrm{d}W}{\mathrm{d}\epsilon}=\frac{\alpha m^2c^4}{\sqrt{3}\pi\hbar^3\omega^2}[\frac{\epsilon^2+\epsilon_f^2}{\epsilon\epsilon_f}K_{2/3}(\xi)+\int_\xi^\infty K_{1/3}(y)\mathrm{d}y]
\label{baier1}
\end{equation}
where $\epsilon_f=\hbar\omega-\epsilon$, $\xi=8u/3\chi$, $u=\gamma^2$, and $\mathrm{d}W=\mathrm{d}w/\mathrm{d}t$ is the probability per unit time. Here $\epsilon$ denotes the energy of one of the produced particles, $\epsilon_f$ that of the other and $\hbar\omega$ the energy of the photon. The functions $K_{1/3}$ and $K_{2/3}$ are the modified Bessel functions of order 1/3 and 2/3, respectively. The Lorentz invariant strong field parameter is given as $\chi=\gamma B/B_0$ where $B_0=m^2c^3/e\hbar=4.41\cdot10^9$ T is the critical magnetic field (corresponding to the electric field ${\mathcal E}_0=1.32\cdot10^{16}$ V/cm) and $\gamma$ is understood here as $\hbar k_\perp/mc$, i.e.\ the photon momentum transverse to the magnetic field, in units of $mc$. The `strong field regime' is reached when $\chi$ becomes comparable to or exceeds 1 - but already at $\chi\simeq0.1$ strong field effects become significant. Thus, for sufficiently energetic photons even the intergalactic fields will appear as a strong field and generally speaking the Universe thus will become `opaque' to these photons. 

The differential probability develops a pronounced minimum around $x=\epsilon/\hbar\omega=1/2$ and peaks at $x\simeq1.6/\chi$ for large $\chi$. This - combined with the fact that the strong field radiation emission tends toward the endpoint of the spectrum - means that as the energy increases beyond the region where $\chi$ is of the order 100, the shower develops as essentially one energetic particle or photon, followed by many low energy particles.
This behaviour is reminiscent of the behaviour of the Landau-Pomeranchuk-Migdal (LPM) effect for pair production (see e.g.\ \cite{Klei99}) which also yields a preference for highly asymmetric pairs.

The total pair production probability per unit time is given as \cite{Baie98}
\begin{equation}
W=\frac{\alpha m^2c^4}{6\sqrt{3}\pi\hbar^2\omega}\int_1^\infty \frac{8u+1}{u^{3/2}\sqrt{u-1}}K_{2/3}(\xi)\mathrm{d}u
\label{baier2}
\end{equation}
In the limit $\chi\ll 1$ the probability is exponentially small, $W\simeq3\sqrt{3}\alpha m^2c^4\chi/16\sqrt{2}\hbar^2\omega\exp(8/3\chi)$, and in the limit $\chi\gg 1$ the result is $W\simeq0.38\alpha mc^2 B/B_0\hbar\chi^{1/3}$, i.e.\ the probability actually diminishes once the energy surpasses the domain where $\chi\simeq10$ \cite{Baie98}. Indications of the `saturation' precursory to the diminishment of the corresponding energy loss by radiation emission has been shown experimentally only recently, \cite{Kirs01,Kirs01b}.

The equations (\ref{baier1}) and (\ref{baier2}) are exact expressions (for $B\ll B_0$), but eq.\ (\ref{baier2}) has an `empirical approximation' \cite{Tsai74} which generally forms the basis for the above mentioned calculations of the pair production in the Earth magnetic field. The decisive parameter for the behaviour is $\chi=\gamma B/B_0$ which depends only on the product of the field and the `Lorentz factor' of the photon, $\hbar k_\perp/mc$. The same $\chi$ applies for strong field effects in radiation emission where $\chi=E/mc^2$.

With the acceleration techniques at hand today, the energy of electrons available for laboratory experiments is limited to $\gamma\leq10^6$. This means that \emph{macroscopic} magnetic fields of the order a few thousand Tesla or electric fields of the order $10^{11}$ V/cm are needed to experimentally test the theory. At present this is certainly beyond reach by conventional means. However, an analogous case is seen for the pair production and radiation emission in single crystals and these phenomena enable a verification of the essential expressions, eqs.\ (\ref{baier1}) and (\ref{baier2}).

\section{Strong fields in crystals}

In the following, a brief introduction to strong field effects in crystals is given with particular emphasis on pair production. However, essentially all conclusions and expressions can be applied also to the case of radiation emission and thus apply for the development of electromagnetic showers as well.
For a more complete introduction to strong field effects in crystals, see  \cite{Baie98}, \cite{Sore96}, \cite{Baie89} and \cite{Sore89}.

\subsection{Theory}

\begin{figure}
\begin{center}
\vspace{-5cm}
\mbox{\includegraphics[width=8cm]{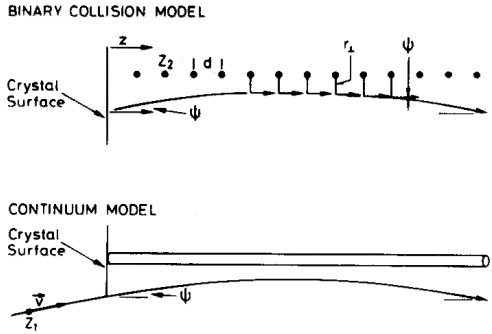}}
\caption[]{\textsl{A schematical drawing of the discrete nature of the scattering centers in a crystal and the resulting continuum approximation.}
\label{fig:scheme}}
\end{center}
\end{figure}

The large fields present near the nuclei in solids may in the case of
single crystals add coherently such that a penetrating charged particle experiences a
continuous field along its direction of motion, see figure \ref{fig:scheme}. If further the particle is
incident with a sufficiently small angle to a particular crystallographic
direction, inside the so-called Lindhard angle, the negatively/positively
charged particle is constrained to
move near to/far from the nuclei and the electron clouds surrounding these. This
is the channeling phenomenon~\cite{Lind65} which has found widespread
applications in physics.
The critical angle for axial channeling is given as
\begin{equation}
\psi_1=\sqrt{4Ze^2/pvd}
\end{equation}
where $Ze$ is the charge of the lattice nuclei and $d$ the spacing of atoms along the axial direction. The critical angle, $\psi_1$, is of the order a few tens of $\mu$rad for e.g.\ 100 GeV electrons. Another well-known example of coherent action is the emission of so-called coherent bremsstrahlung. This appears when crystal planes are crossed at regular intervals \emph{and} at small angles to crystallographic directions during penetration \cite{Pala68}. The small angle requirement is essential for the coherence of the scattering that results in amplification of e.g.\ the radiation spectra.

On the other hand, at high energies (generally above a few GeV for electrons) a seemingly different coherent effect starts to appear. This effect arises due to the continuous field seen by the penetrating particle which acts on an electron with sufficient strength to impose a strongly curved trajectory (as compared to the energy - and thus typical emission angles - of the particle), see figure \ref{fig:scheme}. The curvature - independent of the origin being magnetic or electric fields - gives rise to radiation emission of synchrotron character. Furthermore, this effect persists out to the so-called Baier angle, $\Theta_0$, which is of the order of a few mrad and independent of energy
\begin{equation}
\Theta_0\simeq Ze^2/dmc^2
\end{equation}

The electric fields, ${\mathcal E}$, can be locally up to a few $10^{11}$ V/cm, depending on the crystal type and orientation.
The incident particle moves in these immensely strong fields over distances up to that of the crystal thickness, i.e.\ up to several cm.
For a sufficiently energetic particle the critical parameter $\chi=\gamma{\mathcal E}/{\mathcal E}_0$ can thus reach values near and even beyond unity.
Thereby the behaviour of charged particles in macroscopic strong
fields as \({\mathcal E}_0\) or $B_0$ can be investigated in accelerator based experiments.

It is essentially the same phenomenon which is responsible for the mrad deflection of multi-GeV protons \cite{Baur00} and ions \cite{Ardu97} during the passage of a few cm of a bent crystal. In this case, the equivalent field is as high as a few thousand Tesla.

Strong field effects can be investigated by other means. One example is in heavy ion collisions where the field becomes comparable to the critical field, but the collision is of extremely short duration. Another - technically demanding - example is in multi-GeV electron collisions with terawatt laser pulses where non-linear Compton scattering and so-called Breit-Wheeler pair production are observed \cite{Bamb99}. In none of these cases can the field be considered macroscopic.

\subsection{Experiment}

\begin{figure}
\begin{center}
\hspace{2cm}
\mbox{\includegraphics[width=12cm]{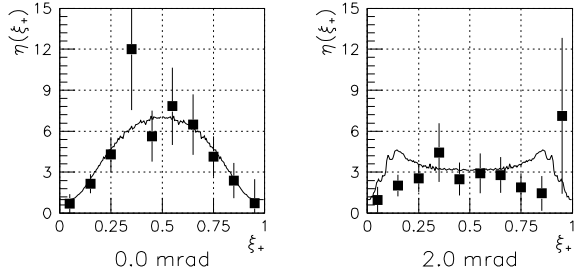}}
\vspace{-0.5cm}
\caption[]{\textsl{Differential enhancement of pair production for photons in the energy interval 80-100 GeV incident around the \(\langle110\rangle\) axis in a germanium crystal. Filled squares are experimental results and the line represents the theoretical values. The results are shown for incidence in the (110) plane at angles 0 and 2 mrad to the \(\langle110\rangle\) axis \cite{Kirs98}.}
\label{fig:ppdiff}}
\end{center}
\end{figure}

In figure \ref{fig:ppdiff} is shown the enhancement (the ratio of radiation lengths for the amorphous and crystalline material) in a germanium  crystal of the pair production probability differential in the energy of one of the produced particles. This is compared with a calculation performed on the basis of eq.\ (\ref{baier1}) for 0 mrad as shown by the line. The calculation is done by assuming a constant field during the formation of the pair and only an average over the fields encountered has to be performed.
For the case of incidence with an angle of 2 mrad to the axis, the line denotes a strong field modified coherent bremsstrahlung calculation \cite{Kono93}.
In both cases a good agreement between experiment and theory is seen, details can be found in \cite{Kirs98}.

\begin{figure}
\begin{center}
\vspace{-0.5cm}
\mbox{\includegraphics[width=7.5cm]{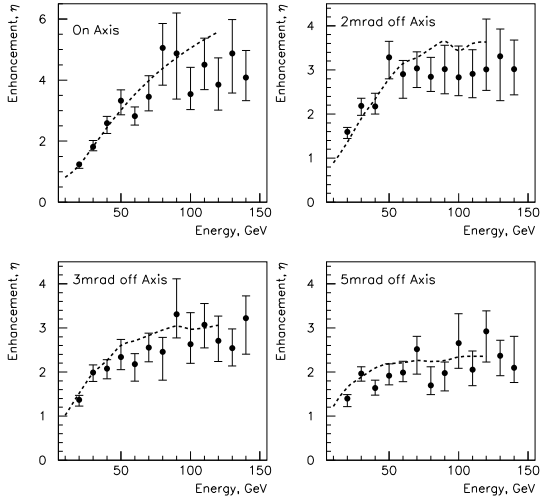}}
\vspace{-4.5cm}
\caption[]{\textsl{Total enhancement of pair production for photons in the energy interval 10-150 GeV incident around the \(\langle100\rangle\) axis in a tungsten crystal. Filled circles are experimental results and the dashed line represents the theoretical values. The results are shown for four angles to the \(\langle100\rangle\) axis as indicated \cite{Moor96}.}
\label{fig:pptot}}
\end{center}
\end{figure}

In figure \ref{fig:pptot} is shown the total enhancement in a tungsten crystal. The relevant averaged values of $\chi$ range up to about 5, i.e.\ strong field effects are expected to be dominant. The theoretical values are based on eq.\ (\ref{baier2}) for the case on axis, whereas for higher angles the lines denote a strong field modified coherent bremsstrahlung calculation \cite{Baie86}. Again a good agreement between experiment and theory is seen, details can be found in \cite{Moor96}.

The slight disagreement at high photon energies seen in figure \ref{fig:pptot} is possibly due to the Landau-Pomeranchuk-Migdal (LPM) effect for pair creation \cite{Kono98} which is not included in the calculation. The LPM effect is not noticeable until photon energies of a few TeV in amorphous materials, but in crystals the multiple Coulomb scattering may increase by orders of magnitude and thus render the LPM effect visible at much lower energies.

\begin{figure}
\begin{center}
\mbox{\includegraphics[width=6cm]{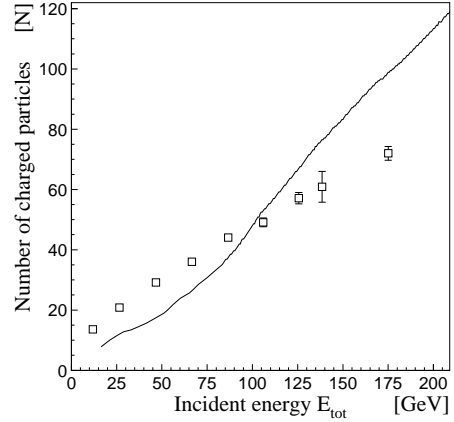}}
\caption[]{\textsl{The number of charged particles exiting an aligned 25 mm thick germanium crystal, compared to a Monte Carlo calculation (continuous line) based on eq.\ (\ref{baier1}) and its equivalent for radiation emission \cite{Baur99}.}
\label{fig:shower}}
\end{center}
\end{figure}

In figure \ref{fig:shower} is shown the number of charged particles exiting an aligned 25 mm thick germanium crystal. A comparison with a Monte Carlo calculation based on eq.\ (\ref{baier1}) \cite{Baie96} shows some disagreement, but in general the trend is correct. For details, see \cite{Baur99}.
It should be noted, however, that for the calculation of radiation emission in crystals a few additional complications arise due to a redistribution of the particle flux inside the crystal and the connected change in e.g.\ multiple Coulomb scattering. It is clear that small inaccuracies in the assessment of the radiation probability or the pair production probability will propagate into amplified discrepancies after a large number of generations in the shower as could be the case for EASs.

\section{Pair production in the geomagnetic field}

\begin{figure}
\begin{center}
\vspace{-0.5cm}
\mbox{\includegraphics[width=8cm]{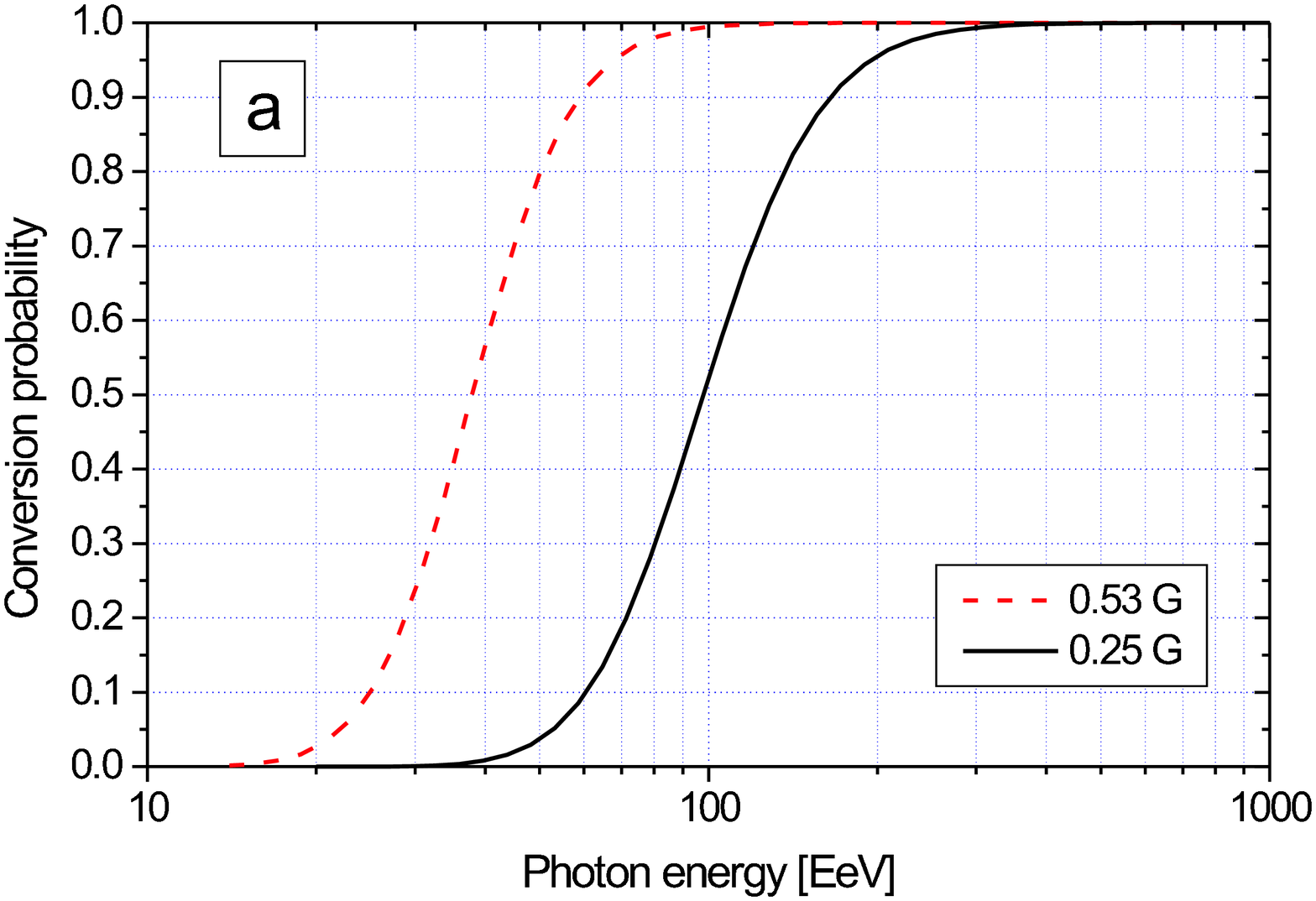}}
\vspace*{-1.0cm}
\mbox{\includegraphics[width=8cm]{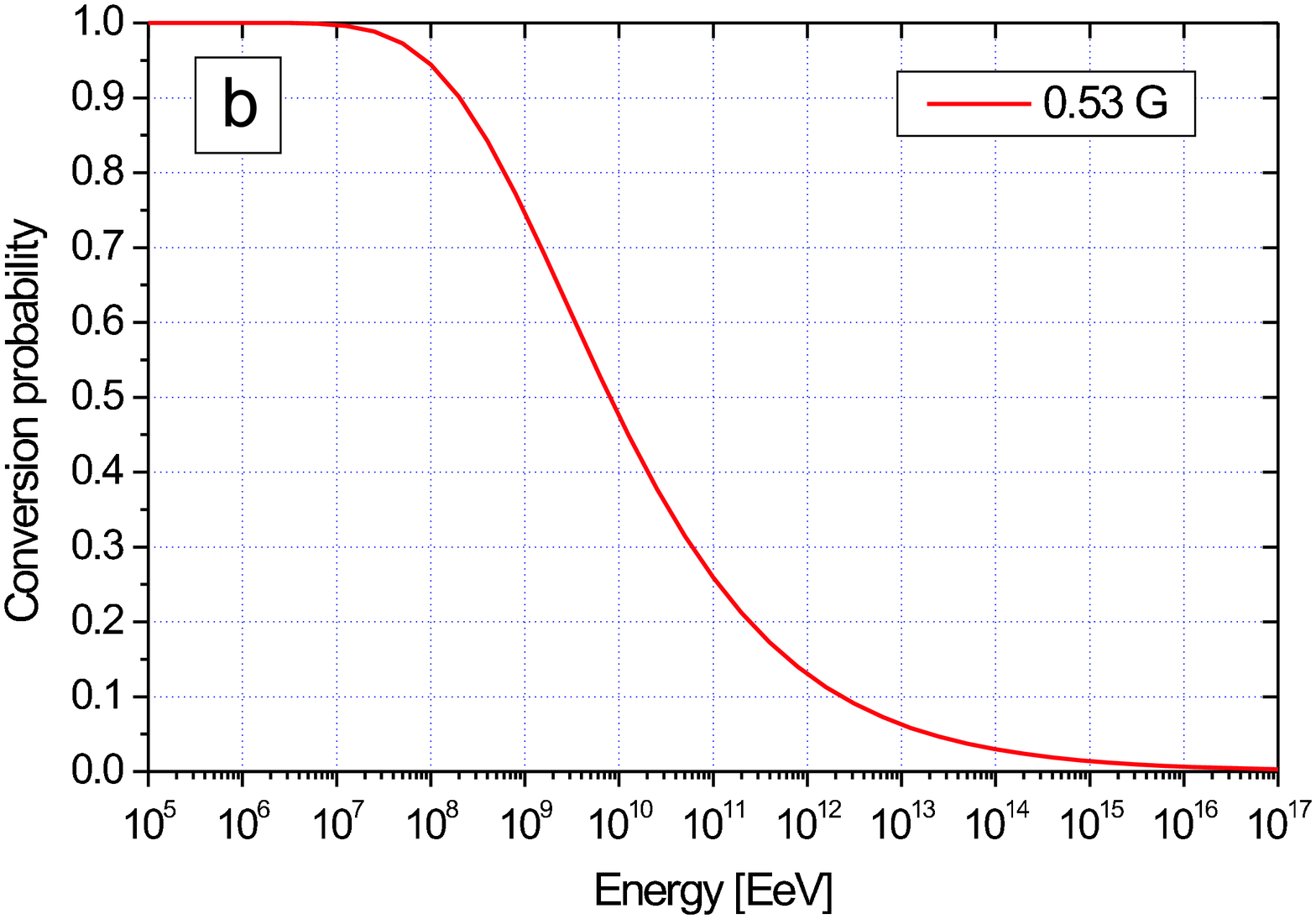}}
\vspace{-0.5cm}
\caption[]{\textsl{The conversion probability of a photon in the geomagnetic field as a function of photon energy for a) energies in the region $10^{1}-10^{3}$ EeV and b) energies in the region $10^{5}-10^{17}$ EeV.}
\label{fig:earthpp}}
\end{center}
\end{figure}

In figure \ref{fig:earthpp} is shown a calculation based on eq.\ (\ref{baier2}) similar to that of McBreen and Lambert \cite{McBr81} or Stanev and Vankov \cite{Stan98} to determine the conversion probability of a photon in the geomagnetic field. A dipole field has been assumed with values of 0.25 G and 0.528 G at the surface of the Earth, corresponding to the two sites of the Pierre Auger Observatory \cite{Cill00}. Figure \ref{fig:earthpp}a reproduces essentially the results of \cite{McBr81,Stan98} for different fields while figure \ref{fig:earthpp}b has been included to show the disappearance of the absorption for sufficently high values of $\chi$. The energy values used in figure \ref{fig:earthpp}b are certainly way beyond what is presently considered reasonable (even beyond the Planck energy) and serve only as an illustration of the $W\propto1/\chi^{1/3}$ behaviour which may be relevant in other cases as discussed in the following.

\section{Pair production in extraterrestrial fields}

\begin{figure}
\begin{center}
\mbox{\includegraphics[width=8cm]{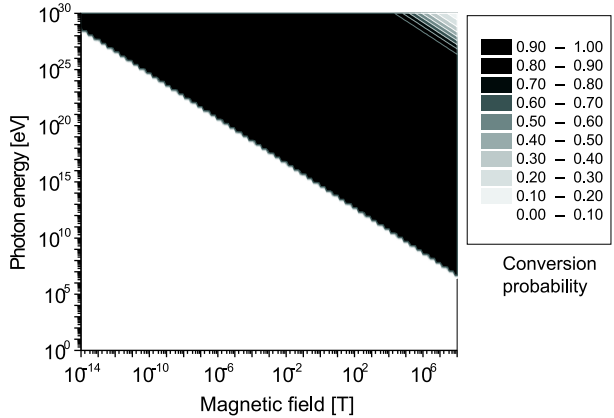}}
\caption[]{\textsl{The conversion probability as a function of photon energy and magnitude of the magnetic field under the assumption implicit in eq.\ (\ref{field1}) with $C=1$.}
\label{fig:absband}}
\end{center}
\end{figure}

Following Hillas \cite{Hill84} there is a suggestive connection between the magnitude of extraterrestrial magnetic fields and their extent. Thus a simplification of the realistic fields can be parametrized roughly by
\begin{equation}
B_{\mu\mathrm{G}}=\frac{C}{L_{\mathrm{pc}}}
\label{field1}
\end{equation}
where $C$ is set here somewhat conservatively to $1$, but can be set to anything within $0.1\leq C\leq10^6$ - in any case the extent and magnitude of many extraterrestrial fields are uncertain \cite{Kron94}. The extent of the field in parsec is given by $L_{\mathrm{pc}}$ and $B_{\mu\mathrm{G}}$ is the field strength in microgauss. By use of eq.\ (\ref{baier2}) the survival probability for a photon traversing regions with magnetic fields can be estimated. The result is shown in figure \ref{fig:absband}. A change in $C$ hardly affects the onset of the absorption band which is why such a large range of accepted values can be replaced by a constant, set to 1 here. The explanation is that the decisive parameter is proportional to $\hbar\omega B$ with an exponential increase, whereas the extent enters linearly. On the other hand, different values of $C$ move the `survival island' in the upper, right corner since in this case there are two 'competing' power-laws.
Thus, for relatively strong fields of limited extent, high energy photons may actually become more penetrating than those of lower energy.

The range of figure \ref{fig:absband} has been set to include the possible value of an `all-pervading' primordial field of $\leq10^{-9}$ G \cite{Kron94} (where clearly the extent is severely underestimated) and to go near the highest field strengths at pulsars of $\simeq10^8-10^9$ T (but respecting $B\ll B_0$).

The above is certainly a rough simplification of the actual extraterrestrial fields which are in reality (as also seen from the `Hillas plot') grouped in small clusters instead of being constrained to the line defined by eq.\ (\ref{field1}). This means that the `absorption band' shown in figure \ref{fig:absband} is an overestimate and should be replaced by `islands' instead. However, this requires a thorough analysis of the extent and magnitude of the actual fields and is beyond the scope of this paper. Furthermore, there are other absorption mechanisms for photons: Collisions with the infrared/optical, cosmic microwave background or universal radio background photons \cite{Bert00}.

\section{Conclusion}

It has been shown in the presented paper that the theoretical approaches used for simulations of photon cascade development in e.g.\ the geomagnetic field are experimentally justified from the point of view of `strong field effects' in crystals.

Actual experiments using crystals may to a large extent simulate the behaviour of shower development in the Earth magnetic field and atmosphere as a number of effects (strong field, LPM, Chudakov as well as e.g.\ ordinary bremsstrahlung) are believed to be present in both cases. In fact, model builders may use data obtained in crystals as test cases for the simulation of many aspects of extended air showers.

\section{Acknowledgments}
It is a pleasure for me to thank Allan H. S{\o}rensen for letting me benefit from his deep insights into 'strong field effects'  through comments that lead to substantial improvements of this paper. Furthermore, I wish to thank the members of the NA43 experiment at CERN who have all made an effort in obtaining the presented experimental data on crystals.

\end{document}